\documentclass{aa}
\usepackage{graphicx}
\usepackage{amssymb }
\begin{document}
   \title{Very low-mass members of the Lupus~3 cloud
     \thanks{Based on observations collected at the European Southern 
       Observatory, La Silla, Chile}
     \thanks{This publication makes use of data products from the Two Micron 
       All Sky Survey (2MASS), a joint product of the University of 
       Massachussets and the Infrared Processing and Analysis centre/California 
       Institute of Technology, funded by the US National Aeronautics and Space 
       Administration and the US National Science Foundation.}
   }

   \subtitle{}

   \author{Bel\'en L\'opez Mart\'{\i},\inst{1}
 	  Jochen Eisl\"offel\inst{2}
	  \and Reinhard Mundt\inst{3}
          }

   \offprints{B. L\'opez Mart\'{\i}}

   \institute{Departament d'Astronomia i Meteorologia, Universitat de Barcelona,
   	Mart\'{\i} i Franqu\`es 1, E-08028 Barcelona, Spain\\
              \email{blopez@am.ub.es}
         \and
	 Th\"uringer Landessternwarte, Sternwarte 5, D-07778 Tautenburg,
   	 Germany\\
              \email{jochen@tls-tautenburg.de}
		\and Max-Planck-Institut f\"ur Astronomie, K\"onigstuhl 17, 
		D-69117 Heidelberg, Germany\\
		\email{mundt@mpia-hd.mpg.de}
            }

   \date{Received ; accepted }

   \abstract{
     We report on a multi-band survey for very low-mass stars and brown dwarfs
     in  the Lupus~3 cloud with the Wide Field Imager (WFI) at the ESO/MPG
     2.2~m telescope on La Silla Observatory (Chile). Our multiband optical
     photometry is combined with available 2MASS $JHK$ photometry to identify
     19 new young stars and 3 brown dwarf candidates as probable members of
     this star forming region. Our objects are mostly clustered around the
     cloud core. Stars and brown dwarfs have similar levels of H$\alpha$
     emission, probably a signature of accretion. One object, a brown dwarf
     candidate, exhibits a near-infrared excess, which may indicate the
     presence of a disk, but its H$\alpha$ emission cannot be confirmed due to
     its faintness in the optical passbands. We also find two visual pairs of
     probable Lupus~3 members that may be wide binaries.

   }

\authorrunning{L\'opez Mart\'{\i} et al.}

   \maketitle

%
   \begin{figure*}[ht]
   \centering
   \includegraphics[width=12cm, angle=-90, bb= 50 70 600 775]{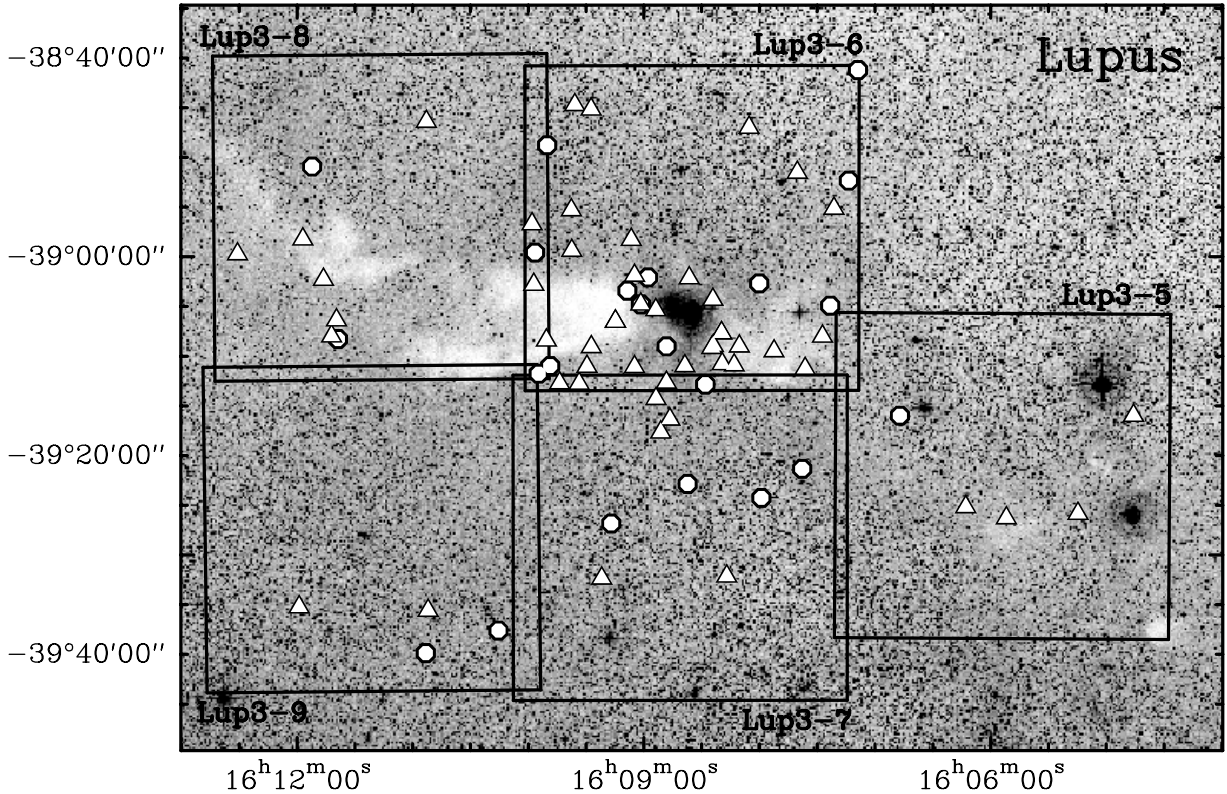}
    \caption{\footnotesize
		Location of our WFI fields (Lup3-5 to 9) in the Lupus~3 cloud.
		The intermediate-mass stars HR~5999 and HR~6000 in the centre
		of the cloud core are seen towards the south in our field
		Lup3-6. The circles and triangles indicate the positions of the
		objects with and without H$\alpha$ emission, respectively. The
		former ones are regarded as probable cloud members.
		}
              \label{fig:lupfield}
    \end{figure*}
%
%
%

\section{Introduction}\label{sec:introdlup}

	In the last years our understanding of the process of brown dwarf
formation has considerably improved. Observations seem not to support the
possibility that these objects form in circumstellar disks, like planets.   
They instead suggest that circum(sub)stellar disks, accretion and mass loss may
be common phenomena in the substellar mass regime (e.g. Muench et al.
\cite{muench01}; Natta et al. \cite{natta02}; Fern\'andez \& Comer\'on
\cite{fernandez01}; Comer\'on et al. \cite{comeron03}; Mohanty et al.
\cite{mohanty03}; Luhman et al. \cite{luhman03}; Scholz \& Eisl\"offel
\cite{scholz04}).  Hence, since young brown dwarfs appear to have similar
properties to low-mass pre-main sequence stars, they may be produced by a
similar mechanism.

	Still, some questions remain open. Theoretical models considering the 
simple collapse of molecular clouds usually fail to reproduce the observed
numbers of substellar objects. It is not clear how the accretion process stops
before the central object becomes massive enough to start the hydrogen burning.
Moreover, the observed star-brown dwarf and brown dwarf-brown dwarf binary
frequencies are difficult to explain with a simple extrapolation of the
low-mass star formation models (e.g. Close et al. \cite{close03}; Bouy et al.
\cite{bouy03}).

	It has been suggested that the characteristics of the environment may
play an important role in the formation of brown dwarfs.  For instance, Lucas
\& Roche (\cite{lucas00}) argue that winds from  the massive stars in
\object{Orion} might have evaporated the gaseous envelopes of the nearby star
forming cores, preventing them from growing over the hydrogen burning mass
limit. This would also explain why few brown dwarfs are known in the
\object{Taurus} star forming region, which does not contain such massive stars
(Brice\~no et al. \cite{briceno98}; Mart\'{\i}n et al. \cite{martin01}).  In
\object{Chamaeleon~I}, though, we do not find any clear relationship between
the brown dwarfs and the intermediate mass stars in the cloud, because they are
also found near a cloud core devoid of optical objects (L\'opez Mart\'{\i} et
al. 2004, hereafter \cite{lm04}). The possibility of formation of free-floating
substellar objects by photo-erosion of protostellar cores has been studied by
Whitworth \& Zinnecker (\cite{whitworth04}), who conclude that this mechanism
should be very efficient in producing large numbers of brown dwarfs, operating
over a wide range of conditions. However, it requires a relatively massive
initial core to form a single low-mass object.

	Reipurth \& Clarke (\cite{reipurth01}) proposed that brown dwarfs are
stellar embryos ejected from multiple systems by dynamical interactions. This
hypothesis is supported by several numerical simulations (e.g. Bate et al.
\cite{bate02}; Delgado Donate et al. \cite{dd03}). In a recent paper, Kroupa \&
Bouvier (\cite{kroupa03}) use available observational results and theoretical
considerations to favour the ejection scenario as the probable origin of the
different brown dwarf populations of Taurus and Orion. However, their analysis
yields a number of brown dwarfs per star that is lower than inferred by
independent measurements of the substellar mass function in the Galactic field.

	To distinguish between these suggested brown dwarf formation scenarios
it is crucial to obtain unbiased estimates of the substellar mass function from
observations of statistically complete and homogeneous populations of very
low-mass stars and brown dwarfs in various regions. Star forming regions are of
particular importance since their youth ensures that their members have not yet
suffered from strong dynamical evolution of the entire region.

	The \object{Lupus} dark cloud complex consists of five molecular
clouds, named Lupus~1 to 5 (Tachihara et al. \cite{tachihara96}), seen as
streams or dust lanes in optical and near-infrared images. At least 65 T~Tauri
stars are known  within the complex (e.g Schwartz \cite{schwartz77}; Krautter
\cite{krautter91}; Hughes et al. \cite{hughes94}), most of them concentrated in
the \object{Lupus~3} cloud. Two A-type stars, \object{HR~5999} and
\object{HR~6000}, are seen towards the centre of this  cloud. Their membership
of  the region is controversial, though, because the  measured distance to
these stars is in disagreement with other distance  estimates to the complex
(e.g. Crawford \cite{crawford00}). 

	A peculiarity of Lupus is that the distribution of spectral types for
the stars in these clouds is dominated by mid\,M-type objects, in contrast to 
other star forming regions like Taurus-Auriga, which contain many late\,K-type 
T~Tauri stars, as noted by many authors (e.g. Appenzeller et al.
\cite{appenzeller83}; Hughes et al. \cite{hughes94}; Wichmann et al.
\cite{wichmann97}). For this reason, Lupus appears as a very promising region
to search for young brown dwarfs. Recently, Comer\'on et al. (\cite{comeron03})
reported a bona-fide brown dwarf, \object{Par~Lup3-1}, from a small H$\alpha$
survey  around the stars HR~5999 and HR~6000. A recent survey in the
near-infrared by Nakajima et al. (\cite{nakajima00}) has also provided some
brown dwarf  candidates.

	In this paper we present the results from our multiband survey for
brown dwarfs in the Lupus~3 cloud. The method followed is the same as in our
previous survey in Chamaeleon~I (\cite{lm04}). The current paper is structured
as follows: Our observations are described in Sect.~\ref{sec:wfilup}. The
candidate selection and brown dwarf identification are explained in 
Sect.~\ref{sec:bdlup}. In Sect.~\ref{sec:distlup} we analyse the observed 
spatial distribution of our objects and the frequency of binaries. Their
H$\alpha$ emission properties are discussed in Sect.~\ref{sec:hallup}. Finally,
Sect.~\ref{sec:lupconcl} summarises our conclusions.

%
\section{Observations}\label{sec:wfilup}

   \begin{table}
   \centering
   \footnotesize
      \caption[]{\footnotesize Log of WFI observations in Lupus~3}
         \label{tab:luplog}
     \vspace{0.5cm}      
         \begin{tabular}{p{0.2\linewidth}l c c c}
            \hline
            \hline
Date & Field & $\alpha$ (2000) & $\delta$ (2000) \\
            \hline
28 May 1999 & Lup3-6 & 16$^h$ 08$^m$ 36.5$^s$ & -38$^{\circ}$ 57$^{\prime}$ 36$\farcs$7 \\
30 May 1999 & Lup3-7 & 16$^h$ 08$^m$ 26.3$^s$ & -39$^{\circ}$ 28$^{\prime}$ 47$\farcs$1 \\
30 May 1999 & Lup3-8 & 16$^h$ 11$^m$ 01.2$^s$ & -38$^{\circ}$ 56$^{\prime}$ 27$\farcs$9 \\
 3 Jun 1999 & Lup3-5 & 16$^h$ 05$^m$ 40.6$^s$ & -39$^{\circ}$ 22$^{\prime}$ 32$\farcs$2 \\
 3 Jun 1999 & Lup3-9 & 16$^h$ 11$^m$ 07.8$^s$ & -39$^{\circ}$ 27$^{\prime}$ 37$\farcs$6 \\
\hline
\end{tabular}
\footnotesize     
   \end{table}

   \begin{table}
   \centering
   \footnotesize
      \caption[]{\footnotesize Exposure times (in seconds) for the images taken 
      		in each selected filter.}
         \label{tab:times}
     $$ 
         \begin{array}{p{0.2\linewidth}c c c c}
            \hline
            \hline
Filter & $Short~exp.$ \ & $Interm.~exp.$ \ & $Long~exp.$ \ \\
            \hline
Rc/162 &  5 & 60 & 600 \\
Ic/lwp &  5 & 30 & 600 \\
Halpha/7  & 15 & $-$ & 600 \\
855/20 &  16 & 300 & 600 \\
915/28 &  8 & 100 & 600 \\
            \hline
         \end{array}
$$
   \end{table}

	An area of about 1.6~${\Box}^{\circ}$ was observed in the Lupus~3 cloud
with the Wide Field Imager (WFI) at the ESO/MPG 2.2~m telescope on La Silla 
Observatory (Chile). The surveyed area corresponds to five WFI fields, each of
$34\arcmin\times33\arcmin$ in size (see Fig.~\ref{fig:lupfield}). Four of the 
fields (Lup3-6, 7, 8 and 9) define a rectangle centred on the cloud core, seen
as an elongated dust lane in our images. The two A-stars HR~5999 and HR~6000 
are placed in the southern part of the field Lup3-6. The Lup3-5 field lies
towards the south-west of this area, also following the orientation of the
core. Table~\ref{tab:luplog} contains the coordinates of the WFI field centres
and the date of observation for each field. The observations were carried out
in the same three nights as our  analogous survey in Chamaeleon~I
(\cite{lm04}).

	We observed in two broad-band filters, R and I, in a narrow-band filter
centred in the H$\alpha$ emission line (H$\alpha$/7), and in two medium-band
filters, M855 (855/20) and M915 (915/28). The last two allow for a photometric
spectral type classification (see Sect.~\ref{sec:sptlup} below). For each field
and filter, except for H$\alpha$, three images with different exposure times
were taken (see Table~\ref{tab:times}). In the case of H$\alpha$, we took only
two different exposures per field. 

	Data reduction, object search, and photometry were performed within the
IRAF environment\footnote{\footnotesize IRAF is distributed by the National
Optical Astronomy Observatory (NOAO), operated by the Association of
Universities for  Research in Astronomy, Inc. under contract to the US National
Science Foundation. }  in the way explained in \cite{lm04}. After the standard
reduction (bias subtraction and flatfield division), all images were divided
through an illumination mask to correct for an irregular illumination pattern.
For the reddest filters (I, M855, and M915), it was also necessary to subtract
a fringe mask to correct for the strong fringing pattern. For each filter, both
masks were created by combination of the science exposures as explained by
Scholz \& Eisl\"offel (\cite{scholz04}).\footnote{\footnotesize  An alternative
method to correct the WFI illumination gradient using a  second-order
polynomial has been proposed by Koch et al. (\cite{koch04}).}

	We ran SExtractor (Bertin \& Arnouts \cite{bertin96}) to produce an
object catalogue of each surveyed field. For the photometry, the DAOPHOT
package was  used (Stetson \cite{stetson87}). PSF photometry was performed in 
order to also measure close pairs of objects. The broad-band photometry was 
calibrated with Landolt (\cite{landolt92}) standard stars. The extinction 
coefficients from the fit of these stars for the R and I bands were also used 
to perform a correction for atmospheric extinction of the H$\alpha$ and the 
M855 and M915 photometry, respectively. No set of standard stars is available
for our narrow and medium-band filters. For a better understanding of the 
emission properties, the H$\alpha$ instrumental magnitudes were then shifted so
that the  main locus of the objects in each surveyed field corresponds to 
H$\alpha$--R$=0$ in a (H$\alpha$, H$\alpha$--R) colour-magnitude diagram (see
Sect.~\ref{sec:bdlup}). The mean error in this shift is of less than 0.01~mag
for the fields Lup3-6, Lup3-7 and Lup3-8, and of about 0.05~mag in the fields
Lup3-5 and Lup3-9. This value is comparable to the mean photometric error (see
below). We estimate that our survey is complete down to R$\simeq$20~mag and
I$\simeq$19~mag. 

	As discussed by Alcal\'a et al. (\cite{alcala02}), the WFI chip-to-chip
photometric variations can be as high as 3\% in the broad-band filters and 5\%
in medium-band filters. This is the main error source in our photometric
calibration, since only a global fit for all the chips could be performed with
the Landolt standard stars. Moreover, a systematic offset was found between
images with different exposure times, in the sense of the resulting
luminosities being brighter (typically by about 0.05~mag) for longer exposure
times. The origin of this offset is so far unknown. To minimise this systematic
error, we determined the offsets between the three different integration times,
and then shifted the short and long integration times to the system of the
medium integration.

	The errors in the completeness range are in general not larger than
5\%. A more detailed discussion of the calibration procedure and the sources of
the  photometric errors can be found in \cite{lm04}. 

	Due to problems with the shutter, several Lup3-7 and Lup3-9 images 
(including the long and intermediate I-band exposures, necessary for the object
selection) exhibit some darker regions at the field edges. Objects in these
regions will not be considered in the following analysis.

%
%
\section{Brown dwarf candidates}\label{sec:bdlup}

\subsection{Candidate selection} \label{sec:candlup}

	In Paper 1, candidate members of Chamaeleon~I were selected around an
empirical isochrone defined by the previously known brown dwarfs and very
low-mass stars in the cloud in a (I, R--I) colour-magnitude diagram. In
Lupus~3, however, there is no sample of known brown dwarfs that could be used
for this purpose. Hughes et al. (\cite{hughes94}) provide broad-band $RIJHK$
photometry of a sample of known T~Tauri stars in the cloud, but most of these
bright objects are saturated in our images. Therefore, we had to rely on the
theoretical models in this case. Candidate members of Lupus~3 were thus
selected around an isochrone  from the models of Chabrier et al.
(\cite{chabrier00}). To do this, we needed an estimation of both the extinction
and the distance to this star forming region. 

	Hughes et al. (\cite{hughes94}) estimated the individual extinction
values towards the Lupus stars in their study. We used these data to derive an
average extinction of A$_{\mathrm{V}}$=0.86$^{+1.35}_{-0.61}$ in our surveyed
region. With this value, and using the following relations:

\begin{center}
\begin{equation}\label{eq:ext1}
\mathrm{A}_{\mathrm{I}}/\mathrm{A}_{\mathrm{V}}=0.482
\end{equation}
\begin{equation}\label{eq:ext2}
\mathrm{A}_{\mathrm{I}}=1.812 \cdot [(\mathrm{R-I})-(\mathrm{R-I})_0],
\end{equation}
\end{center}

\noindent
derived from the extinction law of Rieke \& Lebofsky (\cite{rieke85}), an
average colour excess of

\begin{equation}\label{eq:colxs}
(\mathrm{R-I})-(\mathrm{R-I})_0=0.23^{+0.36}_{-0.16} 
\end{equation}

\noindent
was computed for the Lupus objects. It must be noted here that Rieke \&
Lebofsky's work uses the Johnson R and I filters, while the system used in this
paper is closer to the Cousins one. However, tests using the extinction law of
Mathis (\cite{mathis90}) showed that the sample of selected objects would not
be affected by the use of one or the other extinction law.

	There has been some debate about the actual distance to the Lupus
clouds. Values between 130 and 220~pc can be found in the literature (see e.g. 
Crawford \cite{crawford00} for a discussion). Since our extinction estimate is
based on the work by Hughes et al. (\cite{hughes94}), we will consider, for
consistency, the distance given by these authors, 140$\pm$20~pc. Within the
quoted errors, this value is consistent with most of the more recent
measurements (see e.g. Crawford \cite{crawford00}; Comer\'on et al.
\cite{comeron03}). However, as discussed later in this paper, this might be an
underestimation of the true distance to Lupus~3 (see Sect.~\ref{sec:sptlup}
below).

%
   \begin{figure}[t]
   \centering
   \includegraphics[width=6.5cm, angle=-90, bb= 50 70 600 775]{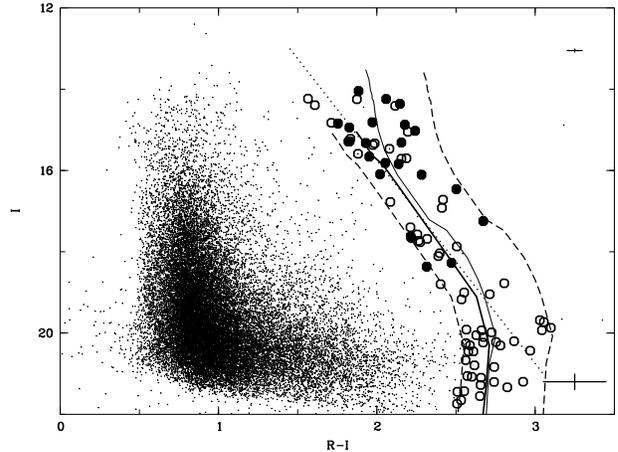}
      \caption{\footnotesize
	(I, R--I) colour-magnitude diagram for our surveyed field in Lupus~3.
	For clarity, only a random selection of the background objects has been
	plotted. Brown dwarf candidates are selected around two isochrones from
	the models of Chabrier et al. (\cite{chabrier00}) corresponding to the
	ages of 1 and 5~Myr (thin and thick solid line, respectively). Objects
	with and without detected H$\alpha$ emission are marked with solid and
	open circles, respectively. The dashed lines indicate the range in
	colour around the isochrones taken as limit for the candidate
	selection, while the dotted line is the empirical isochrone used for
	candidate selection in Chamaeleon~I. The crosses indicate the average
	errors. Note that objects fainter than I$=$20.0~mag (corresponding to
	planetary masses) are not considered in the candidate selection. The
	theoretical models place the stellar/substellar boundary at
	I$\simeq14$~mag for the distance, extinction, and age interval
	considered.} 
	\label{fig:rilup}
   \end{figure}
%

%
  \begin{figure}[t]
   \centering
   \includegraphics[width=6.5cm, angle=-90, bb= 50 70 600 775]{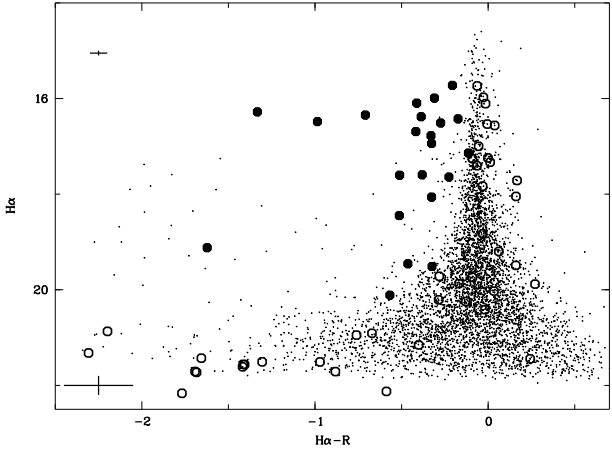}
      \caption{\footnotesize
	(H$\alpha$, H$\alpha$--R) colour-magnitude diagram for our surveyed
	field in Lupus~3. For clarity, only a random selection of the
	background objects has been plotted. Our candidates with and without 
	H$\alpha$ emission are plotted as solid and open circles, respectively.
	H$\alpha$ emitters are placed in the left part of the diagram, having
	negative H$\alpha$--R colour index. (Note that objects not visible in
	our H$\alpha$ images could not be plotted.)} 
      \label{fig:harlup}
   \end{figure}

	Fig.~\ref{fig:rilup} shows the (I, R--I) colour-magnitude diagram for
our surveyed field in Lupus~3. The solid lines are a 1~Myr and a 5~Myr
isochrone from the Chabrier et al. (2000) models, shifted to the estimated
distance and extinction of the cloud. These isochrones span the age range of
the Lupus stars. The dashed lines indicate the position of the isochrones
according to the upper and lower limits for the extinction derived in the above
calculation. A large number of objects are found between these two lines. Note
that there are very few objects to the right of this area; the density of
objects is also clearly diminished left of our search area  with respect to the
main locus of objects seen in the field. Hence, our extinction estimate appears
as a reasonable approximation.

\begin{figure}[t]
   \centering
  \includegraphics[width=6.5cm, angle=-90, bb= 50 70 600 775]{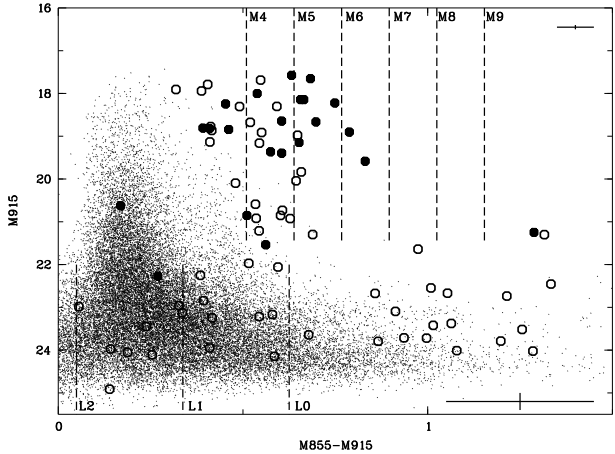}
      \caption{\footnotesize
	(M915, M855--M915) colour-magnitude diagram for our surveyed field in
	the Lupus~3 cloud. Objects with and without H$\alpha$ emission are
	marked with solid and open circles, respectively. Our scale for the
	identification of the spectral type is also indicated. The faintest
	objects in the diagram are regarded as possible L-type objects.}
         \label{fig:mmlup}
   \end{figure}

	In the Lup3-5 field, objects on top of the third CCD chip have bad
flux measurements in the long R exposure due to a badly subtracted background.
The consequence was that these objects had too faint R magnitudes and thus
appeared very red in the (I, R--I) diagram and very blue in the (H$\alpha$, 
H$\alpha$--R) diagram. Visual inspection showed, however, that these objects 
were similarly bright in R and I, and thus very unlikely cloud members. 
Therefore, they were rejected from our candidate list, and are not included in
the colour-magnitude diagrams of Figs.~\ref{fig:rilup}, \ref{fig:harlup} and
\ref{fig:mmlup}. Excluding these objects, we have 85 candidates. 	

	We can compare the diagram in Fig.~\ref{fig:rilup} with the (I, R--I)
colour-magnitude diagram for Chamaeleon~I (Fig.~2 in \cite{lm04}). The dotted
line in Fig.~\ref{fig:rilup} indicates the position of the empirical isochrone
used for candidate selection in \cite{lm04}. Not unexpectedly, given that both
regions are at similar distances, this line is almost coincident with the 1~Myr
isochrone, except for the faintest objects, for which the R--I colour saturates
according to the theoretical models. However, as discussed in detail in
\cite{lm04}, the uncertainties in the colours for the faintest objects are
large, due both to the photometric errors and to the lack of standard stars
that are red enough for a good photometry calibration in this range. This fact
may have implications for the contamination with background objects among our
faintest candidates, as discussed below.

	All the found objects are fainter than I$\sim$14~mag. According to the
Chabrier et al. (\cite{chabrier00}) models, this magnitude roughly corresponds
to the stellar/substellar boundary (M$\simeq$0.075~M$_{\odot}$) at the assumed
age ($\sim$2~Myr, see e.g. Wichmann et al. \cite{wichmann97}) and distance of
Lupus~3 (140~pc). The effective temperature given by these models for objects
with this mass at this age is around 2900-3000~K, corresponding to spectral
types around M6 according to the scale of Luhman (\cite{luhman99}).  Thus, if
these candidates are indeed members of the star forming region, they  must lie
close or beyond the star/brown dwarf transition. 

	Since Lupus~3 is a nearby dark cloud, we do not expect, in principle, a
high contamination from foreground and background objects. Simulations of the
Galaxy population using the Besan{\c c}on models (Robin \& Crez\'e
\cite{robin86}; Robin et al. \cite{robin03}) towards the direction of the Lupus
complex yield between 6 and 15 objects in the region of interest of the (I,
R--I) colour-magnitude diagram. Taking the most conservative result, we would
expect that only around 15\% of our candidates should be non-members of the
star forming region. However, the colour uncertainties for our faintest objects
mentioned above very probably increase this percentage at the lower magnitude
limit of our survey. Given the large errors for the faintest objects, we
stopped our candidate selection at I$\simeq20$~mag in Fig.~\ref{fig:rilup}.
Since the Chabrier et al. (\cite{chabrier00}) models place the planetary mass
limit at  around this value for our estimated age, distance, and extinction, we
do not expect to have planetary mass objects among our candidates.

	To confirm the youth of our objects, we tested their H$\alpha$
emission. As shown in \cite{lm04}, there is a relation between the H$\alpha$--R
colour index and the logarithm of the equivalent width of the H$\alpha$
emission line, $\log$\,EW(H$\alpha$): In general, the bluer (i.e., more
negative) the H$\alpha$--R colour, the stronger the H$\alpha$ emission.
Observations of previously known brown dwarf candidates in Chamaeleon~I showed
that objects with H$\alpha$ equivalent widths of 11\AA \ already present
H$\alpha$--R colours bluer than the main locus of objects in the field (see
\cite{lm04} for details).

	Fig.~\ref{fig:harlup} shows the (H$\alpha$, H$\alpha$--R)
colour-magnitude diagram for our Lupus survey. Many of the brightest objects
exhibit clear H$\alpha$ emission, i.e. negative values of the H$\alpha$--R
colour index. Some objects are clear non-emitters, having positive values of
the H$\alpha$--R colour index. A number of the faintest candidates also seem to
have H$\alpha$ emission, but this is very uncertain due to the large
photometric errors; therefore, except for an object with an apparent
near-infrared excess (see Sect.~\ref{sec:nirlup} below), they were not further
considered. Moreover, twenty-two of the faintest objects are not detected in
our H$\alpha$ images; hence, they do not appear in Fig.~\ref{fig:harlup}. 
Since nothing can be said about their emitting properties, these 22 faintest
objects were also excluded from our study.

	For the brightest objects in Fig.~\ref{fig:harlup}, we applied the same
criterion as Lamm et al. (\cite{lamm05}), retaining the objects with

\begin{equation}\label{halcrit}
\Delta(\mathrm{H}\alpha-\mathrm{R})=
(\mathrm{H}\alpha-\mathrm{R})_{\mathrm{object}}
-(\mathrm{H}\alpha-\mathrm{R})_{\mathrm{locus}}
<-0.1 
\end{equation}

\noindent
as probable cloud members. We discarded all the faintest candidates (H$\alpha
>$20.5~mag), because the large errors at the lower part of the colour-magnitude
diagram do not allow us to discriminate between emitters and non-emitters.

	Our final sample contains 22 stars and brown dwarf candidates. Note
that more than 65\% of the objects detected in our H$\alpha$ images do not seem
to have H$\alpha$ emission according to this criterion.  Such a low fraction of
objects with H$\alpha$ emission is not surprising. In \object{NGC~2264}, which
has an age (1-3~Myr) very similar to Lupus~3, Lamm et al. (\cite{lamm05}) found
only 18\% of strong H$\alpha$ emitters among their low-mass stars with masses
$M<0.25M_{\cdot}$ On the other hand, some objects might have H$\alpha$ emission
levels too low to be reliably detected with our data. Moreover, not all young
objects show H$\alpha$ emission.  The  scarcity of H$\alpha$ emitters among
very low-mass stars in other star forming regions suggests that many of the
remaining objects from our initial candidate sample belong to the star forming
region. Hence, future observations are likely to increase the number of very
low-mass members in Lupus~3.

%
%
%
%

\subsection{Spectral types}\label{sec:sptlup}

	Our medium-band filter M855 lies in a spectral region including the TiO
and VO absorption bands used to classify M-type objects (see Kirkpatrick et al.
\cite{kirkpatrick91}). The filter M915 covers a pseudocontinuum region in these
late-type objects. In \cite{lm04}, we showed that there is a relation between
the M855--M915 colour index and the M-subspectral type using synthetic
photometry from the models of Baraffe et al. (\cite{baraffe98}). In the range
M4-M9, a linear fit is a good approximation of this relation. The corresponding
correlation coefficients were determined  using photometry of known objects of
late spectral type. On the other hand, for spectral types earlier than M4 the
M855--M915 colour saturates, because the TiO and VO features are not prominent
or not present at all in the spectra. A second correlation, of opposite sign,
was found for spectral types L0-L2, due to the progressive disappearance of the
TiO and VO from the spectra. In this case, however, the uncertainties are
larger than for late\,M-type objects, given that extincted early\,M-type
objects would be placed in the same region of the colour-magnitude diagram. 

	This method enabled us to derive spectral types for our brown dwarf
candidates from their position in a (M915, M855--M915) colour-magnitude
diagram. We estimated an error of 1-2 subclasses in this calibration. A
detailed discussion of this photometric spectral type classification, its error
sources, and its limitations can be found in \cite{lm04}.

	Fig.~\ref{fig:mmlup} shows the corresponding (M915, M855--M915)
colour-magnitude diagram. The vertical lines indicate our scale for the
spectral type determination. It is evident from this figure that most of our
objects have spectral types M4-M5, thus lying above, but close to, the
substellar boundary. A few objects have spectral types M6 or later, being brown
dwarfs or brown dwarf candidates. A ``turning back'' towards the left, which
could be  indicative of the beginning of the L spectral sequence, is seen for
the faintest objects. This change of tendency happens at a similar M915
magnitude as in the case of Chamaeleon~I (Fig.~8 in Paper 1), as expected if
both clouds have similar age, distance, and average extinction. 

	About one half of the fainter candidates are found at the lower left
part of the diagram, in the main locus of the objects seen in the surveyed
field. In this group, only Lup~654 and Lup~706 seem to have H$\alpha$ emission.
The other two or three objects in the lower part of the diagram do not show
clear H$\alpha$ emission in Fig.~\ref{fig:harlup}. Except for Lup~642, which
shows a near-infrared excess and apparently no extinction (see below), they
could not be confirmed as cloud members, and are thus rejected. 

	The WFI photometry results, spectral types, and classification for our
22 new Lupus~3 members are summarized in Table~\ref{tab:lupphot}. As in
\cite{lm04}, all objects with spectral types earlier than M5.5 (16 objects) are
classified as very low-mass stars. Four objects lie in the transition from
stars to brown dwarfs with spectral types between M6 and M6.5. Three objects
Lup~642, Lup~654, and Lup~706, have estimated spectral types later than M6.5.
Although they are considered brown dwarf candidates in the subsequent
discussion, it must be kept in mind that their spectral types (all later than
M9) are very uncertain. 

	Note that the largest group of objects corresponds to spectral type M5,
as expected from the characteristics of the Lupus region. As mentioned in
Section~\ref{sec:introdlup}, the Lupus~3 cloud is particularly rich in
mid\,M-type objects. It is remarkable, though, that the new members in 
Table~\ref{tab:lupphot} are in general fainter than expected for their 
estimated spectral types: As stated in the previous section, with the assumed
distance and extinction, an object with I$\simeq$14~mag should be of spectral 
type M6 according to the theoretical models. However, for such objects (see 
e.g. Lup~915 in Table~\ref{tab:lupphot}) we derive a spectral type M4, about 2 
subclasses earlier than the theoretical prediction, although  still marginally
coincident with the expectations (we estimate an error of about 2 subclasses in
our classification). 

	The discrepancy between the luminosities predicted by the theoretical
models and our results  cannot be explained by the photometric errors alone. It
may be a consequence of the underestimation of the average extinction in the
cloud (see Sect.~\ref{sec:candlup}). To obtain the  measured magnitudes,
A$_{\mathrm{V}}$ should be about 4.5~mag, a value not  unusual in star forming
regions. This would also affect  our derived spectral types, because the
objects would appear redder and  fainter, and hence of later spectral type, in
our colour-magnitude diagram (see discussion in \cite{lm04}). However, most of
our objects do not seem to be highly extincted (see Sect.~\ref{sec:nirlup}). It
may also be that the distance to  Lupus~3 has been underestimated. At 200~pc
and with the extinction estimate from Sect.~\ref{sec:bdlup}, a M4 object would
have I$\simeq$13~mag, which is closer to our measured values. Indeed,
comparison of the predictions of the Lyon models (Baraffe et al.
\cite{baraffe98}) with the photometry and spectral types of the mid M-type
stars studied by Hughes et al. (\cite{hughes94}), whose distance value has been
taken in this work, seems to indicate that, if the extinctions given by these
authors are correct, the objects are too faint for their spectral types. For
instance, \object{Sz~112}, a M4 star with an estimated extinction of
A$_{\mathrm{V}}=$0.85~mag, very close to our assumed average value, has
according to Hughes et al. (\cite{hughes94}) an I-magnitude of 13. This also
hints to a wrong distance estimate.

	A further possibility would be that these objects appear underluminous
for their spectral types due to surrounding dust or accretion. Such objects
have been reported in other star forming regions (e.g. Fern\'andez \& Comer\'on
\cite{fernandez01}; Luhman et al. \cite{luhman03}).  Although it seems unlikely
that most of our Lupus members are affected by such underluminousity, only
reported for a minority of objects so far, the possibility cannot be completely
excluded without a safer distance and age determination.

	There are not many previous optical surveys in Lupus~3 that are deep
enough to be compared with our study. We recovered one object from the list
provided by Hughes et al. (\cite{hughes94}), \object{Sz~113} (Lup~609s). Our
derived spectral type for this object (M5) is later but still consistent,
within our estimated errors, with the classification of these authors (M4). On
the other hand, Comer\'on et al. (\cite{comeron03}) estimate an even later
spectral type (M6) for the same object. They suggest that the discrepancy
between their result and that of  Hughes et al. is due to the different
spectral ranges used to perform the  classification. In view of this, we
decided to keep our derived spectral type for this object. The classifications
by other authors are given in a footnote to Table~\ref{tab:lupphot}. Other
objects from Hughes et al. (\cite{hughes94}) seen in our images are too bright
and hence saturated.

	Comer\'on et al. (\cite{comeron03}) reported on four new Lupus~3
members identified in a spectroscopic survey of a small area around HR~5999 and
HR~6000. Of them, one (\object{Par-Lup3-2}) is not present in our images
because it falls in an interchip gap. The other three have R--I colours that
place them at the edges of our selection band in Fig.~\ref{fig:rilup}. While
\object{Par-Lup3-4} has a relatively blue R--I colour that may be caused by its
very high H$\alpha$ emission, Par-Lup3-1 and \object{Par-Lup3-3} are placed to
the red of our objects in the (I, R--I) colour-magnitude diagram. Since these
objects are all placed in a small area around the stars HR~5999 and HR~6000, in
a region of high extinction in the Lupus~3 cloud, this might indicate that we
are missing highly extincted objects with our selection criteria. However, as
commented in Section~\ref{sec:candlup}, there are not many objects towards the
right of our selection band, so this loss should not be very significant.
Moreover, objects with very red R--I colours ($\gtrsim$3~mag) are more likely
to be background stars seen through the dark cloud.

\begin{table*}[t]  
\centering
\scriptsize
     \caption[]{\footnotesize Candidate low-mass members of 
       Lupus~3$^{\mathrm{a}}$$^{\mathrm{b}}$}
         \label{tab:lupphot}
      \vspace{0.3cm}

         \begin{tabular}{p{0.12\linewidth}c c c c c c c l l l}
            \hline
            \hline
Name & $\alpha$ (2000) & $\delta$ (2000) & R & I & H$\alpha$ & M855 & M915 & SpT$^{\mathrm{c}}$ & Classification \\
 & $hh~mm~ss.s$ & $ddd~mm~ss.s$ & mag & mag & mag & mag & mag &  &  \\
            \hline
\object{Lup~504}  & 16 06 47.0 & -39 16 15.8 & 16.77 & 14.94 & 16.38 & 18.54 & 18.00 & M4 & star 	\\
\object{Lup~604s} & 16 08 00.2 & -39 02 59.7 & 16.51 & 14.36 & 16.09 & 18.34 & 17.65 & M5.5 & star 	  \\
\object{Lup~608s} & 16 09 08.5 & -39 03 43.8 & 16.30 & 14.24 & 15.99 & 18.21 & 17.57 & M5 & star \\
\object{Lup~609s}$^{\mathrm{d}}$ & 16 08 57.8 & -39 02 23.6 & 17.05 & 14.88 & 16.34 & 18.81 & 18.14 & M5$^{\mathrm{e}}$ & star \\
\object{Lup~605}  & 16 07 14.0 & -38 52 37.9 & 18.96 & 16.46 & 18.45 & 20.41 & 19.58 & M6.5 & trans. obj.  \\
\object{Lup~607}  & 16 08 28.1 & -39 13 09.6 & 18.39 & 16.10 & 18.06 & 19.94 & 19.36 & M5 & star	   \\
\object{Lup~617}  & 16 08 48.2 & -39 09 20.1 & 17.27 & 15.02 & 16.94 & 18.97 & 18.22 & M6 & trans. obj.    \\
\object{Lup~642}$^{\mathrm{f}}$  & 16 09 01.5 & -39 05 06.1 & 22.6: & 19.9; & 21.7: & 23.0: & 23.0; & L2 & BD cand.	   \\
\object{Lup~648}  & 16 09 48.6 & -39 11 17.6 & 17.11 & 15.28 & 16.69 & 19.25 & 18.64 & M5$^{\mathrm{g}}$ & star  \\
\object{Lup~650}  & 16 09 49.8 & -38 49 04.5 & 19.84 & 17.62 & 19.5; & 21.36 & 20.85 & M4 & star	   \\
\object{Lup~652}  & 16 07 09.5 & -38 41 30.3 & 20.7; & 18.36 & 20.1: & 22.1; & 21.54 & M4.5 & star	   \\
\object{Lup~654}  & 16 07 23.4 & -39 05 13.2 & 19.92 & 17.25 & 19.5; & 22.5; & 22.3; & L1 & BD cand.	  \\
\object{Lup~710s} & 16 09 17.1 & -39 27 09.4 & 17.98 & 15.84 & 17.60 & 19.80 & 19.15 & M5 & star	 \\
\object{Lup~713s} & 16 07 37.7 & -39 21 38.8 & 17.61 & 15.66 & 16.28 & 19.69 & 18.90 & M6 & trans. obj.  \\
\object{Lup~706}  & 16 08 37.3 & -39 23 10.8 & 20.7; & 18.27 & 19.1; & 22.5; & 21.24 & L0 & BD cand.   \\
\object{Lup~707}  & 16 08 28.1 & -39 13 09.7 & 18.11 & 16.09 & 17.60 & 20.00 & 19.39 & M5 & star	   \\
\object{Lup~714}  & 16 07 58.9 & -39 24 34.9 & 16.78 & 14.81 & 16.51 & 18.80 & 18.14 & M5 & star	   \\
\object{Lup~802s} & 16 11 51.2 & -38 51 04.2 & 17.11 & 15.29 & 16.78 & 19.30 & 18.84 & M4 & star	   \\
\object{Lup~810s} & 16 09 54.6 & -39 12 03.4 & 16.60 & 14.84 & 16.42 & 19.20 & 18.81 & $<$M4 & star \\
\object{Lup~818s} & 16 09 56.3 & -38 59 52.1 & 17.42 & 15.25 & 16.60 & 19.34 & 18.57 & M6 & trans. obj. \\
\object{Lup~831s} & 16 11 38.6 & -39 08 27.1 & 17.25 & 15.32 & 17.14 & 19.23 & 18.82 & $<$M4 & star \\
\object{Lup~914}  & 16 10 16.1 & -39 37 53.4 & 17.87 & 15.81 & 17.64 & 20.79 & 20.62 & $<$M4 & star \\
\object{Lup~915}  & 16 10 54.1 & -39 40 07.0 & 15.93 & 14.05 & 15.72 & 18.70 & 18.24 & M4 & star \\  
\hline 
\end{tabular}
\begin{list}{}{}
\item[$^{\mathrm{a}}$] H$\alpha$, M855, and M915 magnitudes are instrumental 
magnitudes.
\item[$^{\mathrm{b}}$] Photometric errors: blank: 0.05~mag; semicolon: 0.1~mag;
colon: 0.2~mag
\item[$^{\mathrm{c}}$] Errors in the spectral types: M4-M6: 2 subclasses; 
M6.5-M9: 1 subclass
\item[$^{\mathrm{d}}$] This object is identified with Sz~113 (Schwartz \cite{schwartz77})
\item[$^{\mathrm{e}}$] Hughes et al. (\cite{hughes94}) and Comer\'on et al. 
(\cite{comeron03}) give a spectral type of M4 and M6, respectively.
\item[$^{\mathrm{f}}$] Near-infrared selected (see text).
\item[$^{\mathrm{g}}$] Ambiguous spectral type (see text).
\end{list}
   \end{table*}

\subsection{Infrared detections}\label{sec:nirlup}

%
   \begin{figure}[t]
   \centering
  \includegraphics[width=6.cm, angle=-90, bb= 50 70 600 775]{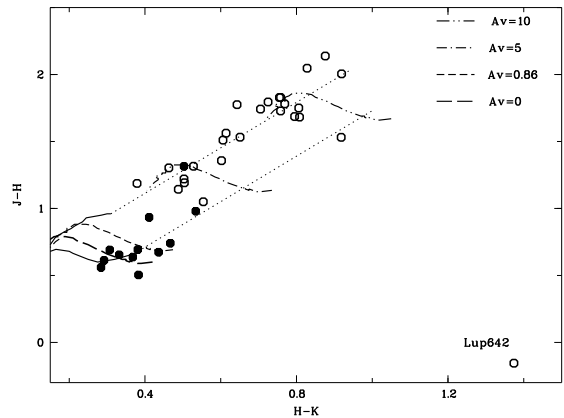}
      \caption{\footnotesize
	Colour-colour diagram for our candidates with near-infrared 
	photometry. Objects with and without H$\alpha$ emission are marked with
	solid and open symbols, respectively. The solid curves indicate the 
	locus of dereddened dwarfs and giants according to Bessell \& Brett 
	(\cite{bessell88}). The dotted lines indicate the direction of the 
	reddening vector up to A$_\mathrm{V}\sim$10~mag. The other lines 
	indicate the position of a 1~Myr isochrone from the models of Baraffe 
	et al. (\cite{baraffe98}) for different extinction values.
      }
         \label{fig:irlup}
   \end{figure}
%

\begin{table}[t]  
\centering
\scriptsize
     \caption[]{\footnotesize 2MASS photometry for our candidate members of Lupus~3}
         \label{tab:irphot}
      \vspace{0.3cm}

         \begin{tabular}{p{0.12\linewidth}c c c c c c c l l l l}
            \hline
            \hline
 Name &     J	  & H	&  K &    2MASS Name   \\
   & mag & mag & mag & \\  
            \hline
 Lup~604s  & 12.145  & 11.452  & 11.071 & 16080017-3902595 \\
 Lup~608s  & 12.202  & 11.642  & 11.358 & 16090850-3903430 \\
 Lup~609s  & 12.465  & 11.724  & 11.257 & 16085780-3902227 \\
 Lup~605   & 14.207  & 13.569  & 13.201 & 16071400-3852375 \\
 Lup~617   & 12.870  & 12.257  & 11.965 & 16084817-3909197 \\
 Lup~642   & 15.381  & 15.536  & 14.162 & 16090155-3905055 \\
 Lup~710s  & 13.548  & 13.044  & 12.661 & 16091713-3927096 \\
 Lup~713s  & 13.240  & 12.566  & 12.130 & 16073773-3921388 \\
 Lup~706   & 15.174  & 14.240  & 13.829 & 16083733-3923109 \\
 Lup~714   & 12.576  & 11.922  & 11.590 & 16075888-3924347 \\
 Lup~802s  & 13.252  & 12.561  & 12.254 & 16115116-3851047 \\
 Lup~810s  & 12.837  & 11.683  & 11.263 & 16095449-3912035 \\
 Lup~818s  & 13.006  & 12.392  & 11.991 & 16095628-3859518 \\
 Lup~831s  & 12.507  & 11.192  & 10.689 & 16113859-3908273 \\
\hline 
\end{tabular}
   \end{table}

	We made use of the database of the 2MASS survey\footnote{\footnotesize
Available online at the URL of the NASA-IPAC Infrared Science Archive (IRSA):
{\tt http://irsa.ipac.caltech.edu/} } to look for near-infrared counterparts of
the 102 objects (with and without H$\alpha$ emission) in our initial candidate
list. In this way, we intended not only to further confirm the youth of our 36
H$\alpha$ emitters, but also to eventually identify new candidate cloud
members. We found 40 positive detections, of which 27 are objects without
detectable H$\alpha$ emission from their H$\alpha$--R colours. All 2MASS
sources are placed at distances not larger than about 1$\arcsec$ from the given
positions for our objects, hence the cross-identifications are very reliable in
all cases.

	Table~\ref{tab:irphot} summarizes the $JHK$ photometry of the new
Lupus~3 members from Table~\ref{tab:lupphot} that were detected by 2MASS.
The photometric errors are in general around 0.02-0.03 in all three bands. In
total, about 50\% of the stars and all the transition objects from
Table~\ref{tab:lupphot} are detected by 2MASS. Two of the brown dwarf
candidates are also found. 

	Above I$\simeq$19~mag, more than 70\% of the  objects were detected by
2MASS, the missing objects in this range being all placed in the outer parts of
the cloud. This is not unexpected, because in those regions the K magnitude for
a given I magnitude should be fainter due to lower reddening. Most of the WFI
candidates not detected in the infrared belong to the faintest optical sources;
hence, they probably lie below the 2MASS detection threshold. 

	Fig.~\ref{fig:irlup} shows the ($J-H$, $H-K$) colour-colour diagram for
our 2MASS objects. Objects with and without H$\alpha$ emission are marked
with different symbols. The solid lines indicate the position of the main
sequence stars according to Bessell \& Brett (\cite{bessell88}). The positions
of a 1~Myr isochrone from the Baraffe et al. (\cite{baraffe98}) models for
several values of extinction are also indicated.

	A degeneracy in the near-infrared colours exists between main sequence 
late-type dwarfs and giants and pre-main sequence low-mass stars. On the other
hand, no matter whether they are true Lupus members or not, the objects in
Fig.~\ref{fig:irlup} are not equally extincted. All the objects without
H$\alpha$ emission have moderate to high values of A$_\mathrm{V}$
($\gtrsim$5~mag). Most of them are found along the reddening line for giants
(upper dotted line); hence, given the lack of H$\alpha$  emission, they are
probably obscured stars seen through or at the edges of the dark cloud. In
contrast, most of the H$\alpha$ emitters are found at low or zero values of
extinction, consistent with our selection criterion from Fig.~\ref{fig:rilup}.
Three objects with H$\alpha$ emission are placed at values of A$_{\mathrm{V}}$
between 2 and 5~mag.

	Only one object, Lup~642, seems to have a near-infrared excess
according to the colour-colour diagram in Fig.~\ref{fig:irlup}. This detection
suggests its belonging to the Lupus~3 cloud, despite not having clear H$\alpha$
emission. However, this object lies very near a previously known Lupus T~Tauri
star (see Sect.~\ref{sec:binlup}), also detected by 2MASS, and its
near-infrared photometry might be contaminated by this brighter source.

	No other object from our initial sample shows excess emission in the
near-infrared. Such an excess is commonly detected in young objects and has its
origin in a circum(sub)stellar disk. It must be noted, however, that the
non-detection of a near-infrared excess does not automatically exclude an
object from having a disk. We will discuss this fact later in
Sect.~\ref{sec:hallup}.

\section{Spatial distribution}\label{sec:distlup}

%
%
   \begin{figure}[t]
   \centering
  \includegraphics[width=6.5cm, angle=-90, bb= 50 70 600 775]{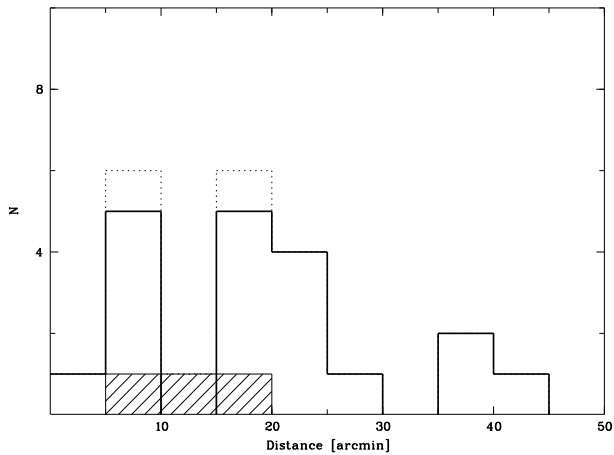}
      \caption{\footnotesize
	Distribution of the low-mass Lupus members with distance to the pair of
	intermediate mass stars HR~5999 and HR~6000. The blank histograms show
	the distribution of the very low-mass stars; the hashed histograms, of
	the brown dwarfs and faint brown dwarf candidates. The common histogram
	for all these objects is showed in dotted lines.}
         \label{fig:objdist}
   \end{figure}
%
%

\subsection{Observed distribution}\label{sec:obsdist}

	Fig.~\ref{fig:lupfield} shows the positions of our new very low-mass
members of the Lupus~3 cloud (dots) and of the objects without H$\alpha$
emission from our initial selection (triangles). Most of the objects are found
within or at the boundaries of the cloud core (seen as an elongated dust lane
in the image), particularly near the  apparent location of the two
intermediate-mass stars HR~5999 and HR~6000. 

	In analogy to the analysis performed for Chamaeleon~I, we have counted
the number of objects at different distance bins from the intermediate-mass
stars; the histograms derived from these counts are shown in
Fig.~\ref{fig:objdist}. Two maxima are found at 5-10$\arcmin$ and
15-20$\arcmin$ from the intermediate mass stars, corresponding to the cluster
of very low-mass objects seen around HR~5999 and HR~6000 in
Fig.~\ref{fig:lupfield}. The three brown dwarf candidates are all placed within
this distance. Further away, the numbers of stars decrease. We remark  that the
bins include all objects at a given distance in all directions. Inspection of
Fig.~\ref{fig:lupfield} shows that, in fact, the objects further than about
15-20$\arcmin$ are not concentred in one particular area. At shorter distances,
however, most of the objects are seen along the dark ridge, a region of higher
density in the cloud (see e.g. Tachihara et al. \cite{tachihara96}). Note also
that there might be some incompleteness in the first bin (distances
0-5$\arcmin$), because very close faint objects might be blended by the
brighter intermediate-mass stars. 

	Although some objects are found far from the cloud core, it is clear
from Figs.~\ref{fig:lupfield} and \ref{fig:objdist} that most of the Lupus
members, and particularly the brown dwarf candidates, are still very close to
their birth site.  This distribution does not seem to support the ejection
model of Reipurth \& Clarke (2001) unless, as argued by Kroupa \& Bouvier
(\cite{kroupa03}), dynamical interactions between members are able to keep the
ejected brown dwarfs within the cluster. Kroupa \& Bouvier (\cite{kroupa03})
successfully explain why a large number of these objects are found in the
star-rich Orion clusters, but not in the more sparse Taurus-Auriga star forming
region. However, Lupus (and also Chamaeleon~I) is an intermediate case between
these two. Detailed calculations for this region should be performed before
stating whether dynamical interactions could have retained the ejected brown
dwarfs also in this cloud. 

	The association with the intermediate mass stars might be accidental,
since their belonging to the cloud is not clear: Several distance measurements
yield a value of about 190-220~pc for these stars (see Wichmann et al.
\cite{wichmann98}; Crawford \cite{crawford00}, and references therein),
inconsistent with other estimations of the distance to the Lupus clouds
(between 140 and 160~pc). On the other hand, in \cite{lm04} no clear
relationship could be established between the very low-mass objects and the
intermediate mass stars in Chamaeleon~I; the former were found near the dark
cloud cores, independently of the presence of higher mass stars in these cores.

\subsection{Wide binaries}\label{sec:binlup}

	To check the existence of multiple systems among our objects, we first
derived the average surface density in our surveyed area. Then we computed the
minimum distance at which an object should have a neighbour if the distribution
were completely uniform. In this way, we estimated a maximum separation of
about 5$\farcs$5 (about 770~AU for a distance of 140~pc) to look for binary
candidates.

	Despite their clustering around the cloud cores, few visual pairs are
found among our probable Lupus members. Only two objects, the transition object
Lup~818s (M6) and the faint brown dwarf candidate Lup~642, are seen near a
previously known Lupus star: The former lies at about 4$\arcsec$ 
($\sim$560~AU) of \object{Sz~119}, while the latter is found at about
6$\arcsec$  ($\sim$840~AU) of \object{Sz~114}. The visual binary frequency of
Lupus~3 is thus very similar to that of Chamaeleon~I, where also only a few
possible systems were seen (\cite{lm04}). 

	Only precise proper motion and radial velocity measurements showing
common space motion may enable us to decide if these pairs are true physical
binaries. It is remarkable, though, that in both Lupus~3 and Chamaeleon~I the
binary candidates have a separation very close to the upper limit separation
estimated for each cloud.\footnote{\footnotesize 
Note that in Chamaeleon~I this cutoff was larger, 12$\arcsec$, due to its
lower object density.
} 
This might indicate that there is no real physical connection between the two 
members of these visual pairs.

\section{H$\alpha$ emission}\label{sec:hallup}

   \begin{figure}
   \centering
  \includegraphics[width=6.5cm, angle=-90, bb= 50 70 600 775]{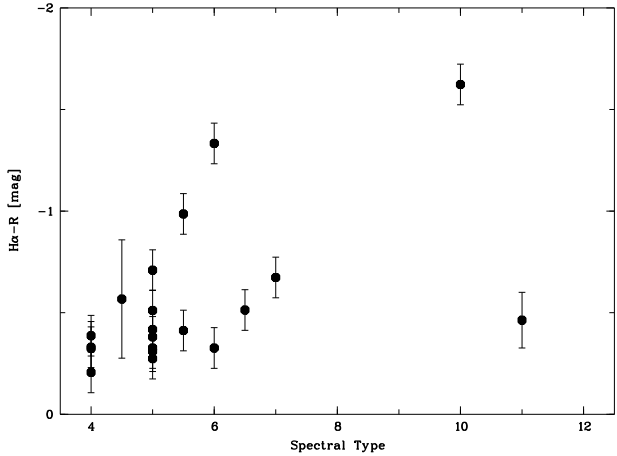}\hfill
      \caption{\footnotesize
		Measured H$\alpha$--R colour index versus  spectral type for
		our new Lupus~3 members. Spectral types 4 to 9 correspond to M4
		to M9, while 10 to 12 correspond to L0 to L2.}
        \label{fig:haspt_lup}
   \end{figure}

   \begin{figure}
   \centering
  \includegraphics[width=6.5cm, angle=-90, bb= 50 70 600 775]{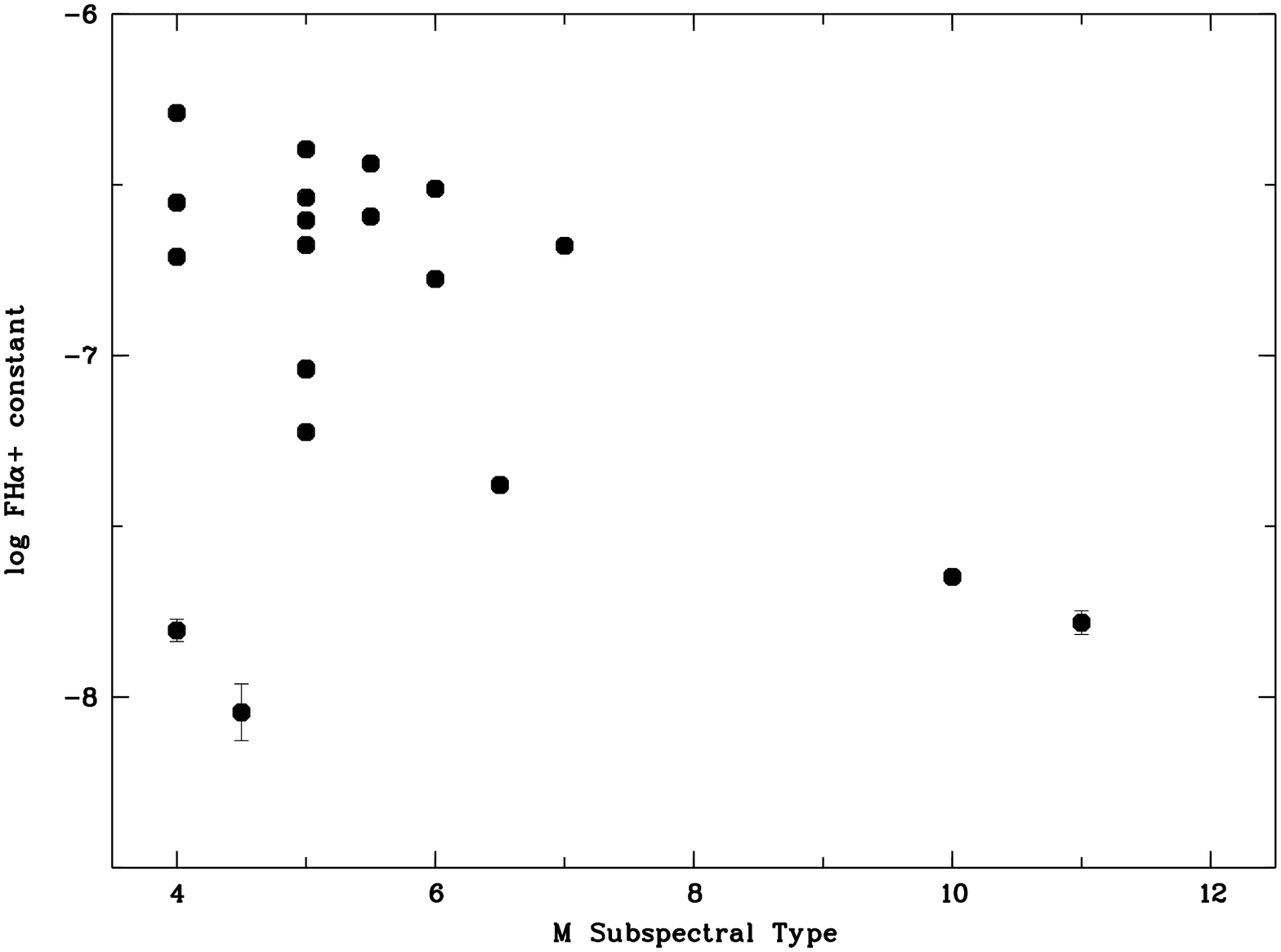}\hfill
  \includegraphics[width=6.5cm, angle=-90, bb= 50 70 600 775]{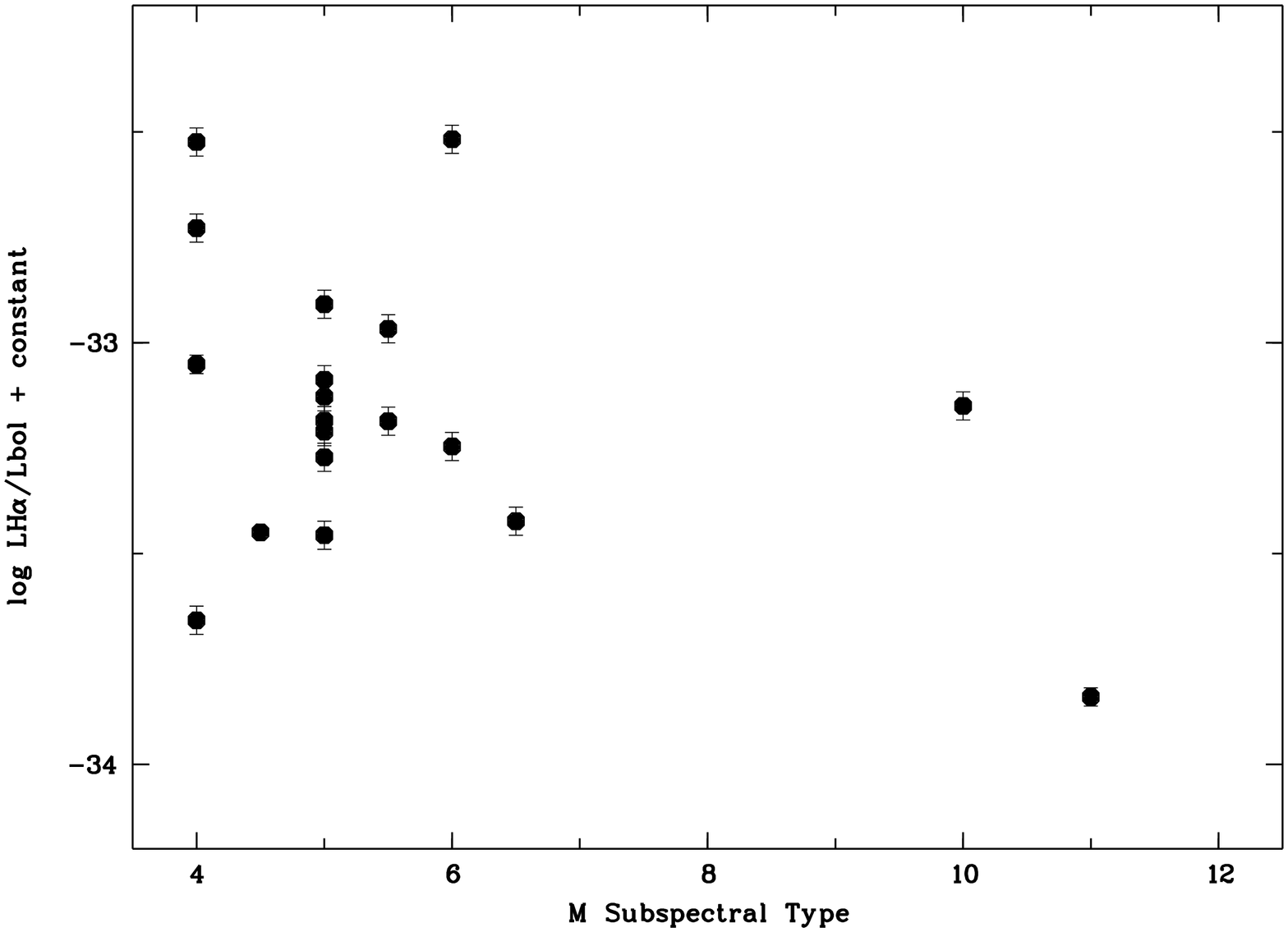}\hfill
      \caption{\footnotesize
		\emph{Upper panel:} Measured H$\alpha$ pseudoflux versus 
		spectral type for our new Lupus~3 members.
       		\emph{Lower panel} Ratio 
		\~L$_\mathrm{H}\alpha$/L$_\mathrm{bol}$ versus spectral type.
		In both panels, spectral types 4 to 9 correspond to M4 to M9,
		while 10 to 12 correspond to L0 to L2.}
        \label{fig:fhaspt_lup}
   \end{figure}

	As explained in Sect.~\ref{sec:bdlup}, the value of the H$\alpha$--R
colour index is related to the equivalent width of the H$\alpha$ line and
hence, to the intensity of the H$\alpha$ emission. The distribution of the
H$\alpha$--R colour with the spectral type is shown in
Fig.~\ref{fig:haspt_lup}. The data seem to indicate an increase of the
H$\alpha$ emission with later spectral type, but the lack of objects of late\,M
spectral type does not allow a definitive conclusion. Even if this were the
case, we caution that the apparent higher emission might just be the effect of
a lower continuum. Moreover, the two possible L-type objects have very
different values of the H$\alpha$--R colour index, which could be an indication
of contamination with earlier type, highly extincted objects. Even with these
uncertainties, it is remarkable that no decrease of the emission (and thus
probably also the accretion rate, as discussed below) is seen across the
substellar boundary.  

	This can be better analysed by directly considering the H$\alpha$ flux
and/or luminosity of our objects. Following \cite{lm04}, we computed a
{\it pseudoflux} \~F$_{\mathrm{H}\alpha}$ as:

\begin{equation}\label{eq:fha}
$\~F$_{\mathrm{H}\alpha}=
\mathrm{F}_{\mathrm{H}\alpha}/\mathrm{F}_0=10^{-m_{\mathrm{H}\alpha}/2.5}
\end{equation} 

\noindent
where $m_{\mathrm{H}\alpha}$ denotes the H$\alpha$ (instrumental) magnitude,
and F$_0$ is a hypothetical H$\alpha$ absolute flux.  We remark that this is a
flux derived from the magnitude measured in our H$\alpha$ band, and not the
flux in the H$\alpha$ spectral emission line, which can only be determined with
spectroscopy. From this quantity \~F$_{\mathrm{H}\alpha}$, a H$\alpha$ 
{\it ``pseudoluminosity''} \~L$_{\mathrm{H}\alpha}$ was computed using the
distance estimate of 140~pc.

	The upper panel in Fig.~\ref{fig:fhaspt_lup} shows the H$\alpha$
pseudoflux as a function of the spectral type for our Lupus objects. The data
apparently indicate a decrease of log~\~F$_{\mathrm{H}\alpha}$ with later
spectral type. This is probably just the expression of the progressive
faintness of our objects. Indeed, if we plot
\~L$_{\mathrm{H}\alpha}/$L$_{\mathrm{bol}}$ instead, which should not be
dependent on the distance or radius of the objects, no significant decrease of
the emission with the spectral type is seen any longer (see lower panel in
Fig.~\ref{fig:fhaspt_lup}). An analogous behaviour was observed also in the
Chamaeleon~I cloud (cf. Fig.~14 in Paper 1). To obtain the bolometric
luminosities, we have followed the method outlined by Comer\'on et al.
(\cite{comeron00}), using a distance value of 140~pc and R--I intrinsic colours
from Kenyon \& Hartmann (\cite{kenyon95}) and Zapatero Osorio et al.
(\cite{zo97}). 

	Since the Lupus members have H$\alpha$ emission properties similar to
the Chamaeleon~I objects, this emission must have a similar origin in both
clouds. In the case of Chamaeleon~I, the strongest emitters were shown to have
a reported mid-infrared excess; hence, their H$\alpha$ emission is most likely
due to accretion from a circum(sub)stellar disk. It seems thus plausible that
also the Lupus objects are accreting from a disk. It is remarkable in this
context that only one of our objects detected by 2MASS may have a near-infrared
excess, a common indication of the presence of a disk around a pre-main
sequence star. Nonetheless, several recent works (e.g. Comer\'on et al.
\cite{comeron00};  Natta \& Testi \cite{natta01}; Natta et al. \cite{natta02};
\cite{lm04}) show that clear near-infrared excess is not always detected in
objects with such cool photospheres. Hence, observations in the mid- and
far-infrared might reveal disks around the Lupus objects. 

If confirmed as an intrinsic property of the objects (and not caused by a wrong
distance or age estimation), a further indication of ongoing accretion would
come from the apparent  underluminosity of our new Lupus members
(Sect.~\ref{sec:sptlup}). Underluminous objects have been reported by other
authors (e.g. Fern\'andez \& Comer\'on \cite{fernandez01}; Luhman et al.
\cite{luhman03}; Comer\'on et al. \cite{comeron03}). They typically show
indications of accretion (and sometimes also mass loss) in their spectra,
particularly a prominent H$\alpha$ emission line. Comer\'on et al.
(\cite{comeron03}) discuss several possible origins for this apparent
underluminosity, such as light scattering by a circum(sub)stellar disk seen
edge-on or extinction through a dusty envelope. They finally suggest that the
evolution of these objects may be critically affected by the accretion process,
thus making them appear fainter than the predictions of the theoretical models,
which are based on too simplified assumptions. Clearly, spectroscopy of our
Lupus objects will be very useful to clarify their nature. 

	Outflows or chromospheric magnetic activity may also contribute to the
observed amount of H$\alpha$ emission from our objects. It is not possible,
however, to disentangle the mechanisms possibly responsible for this emission
with the available data.

\section{Conclusions} \label{sec:lupconcl}

	We have performed a survey for very low-mass objects in the Lupus~3
dark cloud. In an area of about 1.6~$\Box^{\circ}$ we found 19 new stellar and
3 brown dwarf candidate members. 

	Like in Chamaeleon~I, the Lupus low-mass objects are mostly located
near the cloud core, with no evident signs of luminosity segregation. There is
no obvious decrease of the H$\alpha$ emission for the objects with the latest
spectral types. The emission properties of the new Lupus members resemble those
of the Chamaeleon~I members. This indicates a similar origin, most probably
accretion. However, in our analysis of the 2MASS data we find only one object
which could have a near-infrared excess. Nonetheless, given that
circum(sub)stellar disks are not always revealed by near-infrared photometry, 
we expect that many of our objects with H$\alpha$ emission (if not all of 
them) may turn out to have disks in future mid- and far-infrared surveys.

	The most important caveats in our study are, on one side, the
uncertainties in the distance and extinction of the Lupus~3 members, and, on
the other, the uncertainties in our photometric spectral type classification,
especially for the faintest objects in our sample, that might be affected by
extinction. We particularly remark that in general our objects appear too faint
for their estimated spectral type according to the theoretical models. While
the most probable reason for this discrepancy is a wrong distance estimate, we
cannot rule out that the underluminosity is intrinsic to the objects and caused
by the presence of circumstellar disks seen edge-on. It is thus mandatory to
perform spectroscopy to better constrain the true nature of our candidates and
to further study their properties.

\begin{acknowledgements}
We acknowledge much support from I. Baraffe and F. Allard, providing us with
synthetic photometry from their brown dwarf models. We kindly thank
C.~Bailer-Jones for his help with the field selection, and A.~Scholz and
F.~Figueras for useful discussions. J.~E. also thanks E.~Pompei and the
2p2-team for their support during the observations.

We made use of the SIMBAD database, operated at the   
\emph{Centre de Donn\'ees astronomiques de Strasbourg (CDS)} in Strasbourg
(France), and of the NASA/IPAC Infrared Science Archive, which is operated by 
the Jet Propulsion Laboratory, California Institute of Technology, under
contract with the US National Aeronautics and Space Administration.

This work was supported by the German
\emph{Deut\-sche Forschungs\-ge\-mein\-schaft (DFG)}, projects EI\,409/7-1 
and EI\,409/7-2.

\end{acknowledgements}


\end{document}